\documentclass[12pt]{article}
\usepackage{graphicx}
\def\beq{\begin{equation}}
\def\eeq{\end{equation}}
\def\bea{\begin{eqnarray}}
\def\eea{\end{eqnarray}}

\def\roughly#1{\mathrel{\raise.3ex\hbox
{$#1$\kern-.75em\lower1ex\hbox{$\sim$}}}}

\def\sss{\scriptscriptstyle}

\def\bd{B_d^0}
\def\bdbar{{\overline{B_d^0}}}
\def\bs{B_s^0}
\def\bsbar{{\overline{B_s^0}}}
\def\ks{K_{\sss S}}

\newcommand{\fbbs}{f^2_{B_s}\hat{B}_{B_s}}

\newcommand{\delms}{\Delta M_s}
\newcommand{\delmssm}{\Delta M_{s}^{SM}}

\newcommand{\ms}{ M_{12}^s}
\newcommand{\mssm}{ M_{12}^{s,SM}}
\newcommand{\msnp}{ M_{12}^{s,NP}}
\newcommand{\lnps}{H_{NP,S}^{\Delta B=2}}

\newcommand{\lnpa}{H_{NP,A}^{\Delta B=2}}
\newcommand{\lnp}{H_{NP}^{\Delta B=2}}

\newcommand{\heff}{H_{eff}^{\Delta B=1}}
\newcommand{\bsqq}{ b \to s \bar{q} q}
\newcommand{\kpi}{ B^- \to K^- \pi^0}
\newcommand{\phik}{ B_d^0 \to \phi K_s}

\newcommand{\bsbsbar}{$B_s^0$--${\overline{B_s^0}}$}

\def\ApNPqph{{\cal  A}^{\prime,q}  e^{i  \Phi'_q}}  \def\ApNPuph{{\cal
A}^{\prime,u}  e^{i  \Phi'_u}}  \def\ApNPdph{{\cal A}^{\prime,d}  e^{i
\Phi'_d}}   \def\ApNPCqph{{\cal   A}^{\prime   {\sss   C},   q}   e^{i
\Phi_q^{\prime C}}}

\def\bra#1{\left\langle  #1\right|} \def\ket#1{\left| #1\right\rangle}
\def\barpk{{\raise.35ex\hbox  {${\sss  (}$}}--{\raise.35ex\hbox{${\sss
)}$}}}        \def\bbarp{\hbox{$B$\kern-0.9em\raise1.4ex\hbox{\barpk}}}

\def\ApNPqph{{\cal  A}^{\prime,q}  e^{i  \Phi'_q}}  \def\ApNPuph{{\cal
A}^{\prime,u}  e^{i  \Phi'_u}}  \def\ApNPdph{{\cal A}^{\prime,d}  e^{i
\Phi'_d}}   \def\ApNPCqph{{\cal   A}^{\prime   {\sss   C},   q}   e^{i
\Phi_q^{\prime C}}}

   \def\bd{B_d^0} \def\bs{B_s}
\def\bdbar{{\bar   B}_d^0}   
    \def\ks{K_{\sss    S}}

  \def\ApNPqph{{\cal   A}^{\prime,q}  e^{i
\Phi'_q}}    \def\ApNPuph{{\cal     A}^{\prime,u}    e^{i    \Phi'_u}}
\def\ApNPdph{{\cal  A}^{\prime,d}  e^{i \Phi'_d}}  \def\ApNPCqph{{\cal
A}^{\prime {\sss  C}, q} e^{i  \Phi_q^{\prime C}}}      
                  
\def\bra#1{\left\langle  #1\right|} \def\ket#1{\left| #1\right\rangle}

  \def\rr2{{1\over\sqrt{2}}}

\def\.{\!\cdot\!}    \def\:{\cdots}   \def\[{\left[}   \def\]{\right]}
\def\({\left(} \def\){\right)} 
\begin{document}

\begin{flushright} 
\end{flushright}

\begin{center}
\bigskip {\Large  \bf $B_s$ Mixing and New Physics in hadronic  
$b \to s \bar{q} q$
transitions\\}  \bigskip   {\large  Alakabha  Datta
$^{a}$\footnote{datta@physics.utoronto.ca} }
\end{center}

\begin{flushleft}
~~~~~~~~~~~$a$: {\it Department  of Physics, University of Toronto,}\\
~~~~~~~~~~~~~~~{\it  60 St.\  George Street,  Toronto, ON,  Canada M5S
1A7}\\
\end{flushleft}

\begin{center} 
\bigskip (\today) \vskip0.5cm {\Large Abstract\\} \vskip3truemm
\parbox[t]{\textwidth} {  
We study the implication of the recent $\delms$ measurement on $\bsqq$ transitions. We show that it is possible, in the presence of a flavour symmetry, that the phase in $\bs$ mixing may be unobservable even with new CP odd phases in $b \to s$ transitions. These phases may then produce new CP odd effects in certain $ \bsqq$ transitions like $ B \to K \pi$ but not in others like $ B_d \to \phi K_s$. Working in a two higgs doublet model of new physics we discuss  the allowed NP contribution to $B \to K \pi$ and
 $ B_d \to \phi K_s$ decays  with the new $\delms$ measurement.

  }
\end{center}

\thispagestyle{empty} \newpage \setcounter{page}{1}
\baselineskip=14pt

\section{Introduction}
 The   $B^0_s-\bar B^0_s$ mass difference, recently measured by
 the D\O~\cite{D0} and CDF~\cite{CDF} collaborations , is given by,
\bea
\label{Bsmix}
&& 17\,{\rm ps}^{-1} < \delms < 21\,{\rm ps}^{-1} 
  ~~~\mbox{(90\%\,CL,~ D\O)}\,, \nonumber\\
&& \delms = (17.33 ^{+0.42}_{-0.21} \pm 0.07)\, {\rm ps}^{-1} ~~~\mbox{(CDF)}\.
\eea
This result 
is consistent with the SM predictions, which are estimated 
as $21.3 \pm 2.6$ ps$^{-1}$ by the UTfit group~\cite{utfit} 
and $20.9^{+4.5}_{-4.2}$ ps$^{-1}$ 
$21.7^{+5.9}_{-4.2}$ ps$^{-1}$
by the CKMfitter group~\cite{ckmfitter}. The implication of this measurement have been studied in several recent papers \cite{model, zoltan,ball}.
 Given the hadronic uncertainties in the SM prediction, along with additional  uncertainties when NP is included, the present measurement of $\delms$ does not provide a strong constraint on 
NP\cite{zoltan, ball}. 

A measurement of a large phase in $B_s$ mixing will be a definite signal of NP as the SM prediction for this phase is $ 1.045^{+0.061}_{-0.057}$ degrees 
\cite{bsphase} and is tiny. NP phases should then also appear in $ \Delta B=1$, $\bsqq $($q=u,d,s$) hadronic transitions. However a measurement of no significant NP phase in $\bs$ mixing  will not necessarily rule out new NP weak phases in $\bsqq $ transitions. In fact Ref.~\cite{nir} finds that combining the measurements of the mass difference $\delms$ and that of the semileptonic CP asymmetry one can constrain the $\bs$ mixing to be small. 
In this paper we show that one can construct a flavour symmetry  that causes new CP odd phase, from the $b \to s$ vertex,  in $\bs$ mixing to be unobservable  but allows new CP odd effects in $\Delta B=1$, $\bsqq$ transitions. 

There is now a lot of data on rare FCNC hadronic decays from various experiments. The data, specially involving $\bsqq $ transitions, have been somewhat difficult to understand within the standard model(SM). First,   within the SM, the measurement of
the CP  phase $ \sin {2 \beta}$  in $\bd(t) \to J/\psi  \ks$ should be
approximately  equal to that  in decays  dominated by  the quark-level
penguin transition $ b \to s q{\bar q}$ ($q=u,d,s$) like $\bd(t) \to \phi
K_s$,  $\bd(t) \to \eta^{\prime}  K_s$, $\bd(t)  \to \pi^0  K^0$, etc.
However, there is  a difference between the measurements  of $ \sin {2
\beta}$ in the $ b \to s $ penguin dominated modes ($ \sin {2 \beta} =
0.50 \pm 0.06$) and that in  $\bd(t) \to J/\psi \ks$($ \sin {2 \beta}=
0.685 \pm 0.032)$ \cite{ Babar, Belle,hfag}. Note that the
$ \sin {2 \beta}$ number for the $ b \to s $ penguin dominated modes
is the average of several modes.  The effect of new physics
can be different for different final hadronic states and so the
individual $ \sin {2 \beta}$ measurements for the different modes
are important.
Second, the  latest data on  $B\to\pi K$ decays (branching  ratios and
various  CP  asymmetries)  appear  to  be  inconsistent  with  the  SM
\cite{BKpidecays1,BKpidecays2a}.  {\footnote{ A cleaner test of the SM
could be provided by looking at the quasi-exclusive decays $B \to K X$
\cite{dattaKX} rather  than the  exclusive $B \to  K \pi$  decays.}  }
Third, within the SM, one  expects no triple-product asymmetries in $B
\to \phi K^*$  \cite{BVVTP}, but BaBar has measured  such an effect at
$1.7\sigma$ level  \cite{BaBarTP}.  There  are  also polarization  anomalies
where the decays $\bdbar \to \phi K^*$ and $B^- \to \rho^- K^*$ appear
to  have large  transverse  polarization amplitudes  in conflict  with
naive SM expectations \cite{hfag, phiKstarexp,rhoKstarexp,jsmith,rhoKstar}.
While  these  deviations certainly  do  not  unambiguously signal  new
physics(NP), they give reason to speculate about NP explanations of the
experimental data. Hence, it is interesting to study the implication of the measurement of $\delms$ for $\bsqq$ transitions. For instance, one could check if the NP fits to the $B \to K \pi$ decays presented in Ref.~\cite{BKpidecays2a} are consistent with $\delms$ measurements. 

The paper is organized in the following manner: In the next section we study NP phases in $\bs$ mixing and a flavour symmetry under which the new phase in $\bs$ mixing  remains hidden even with NP with non trivial weak phases. We then discuss how these  new weak phases affect hadronic $\bsqq$ transitions.
In the following section we present details of a two higgs model already presented in Ref.~\cite{23} to demonstrate explicitly some of the general ideas of sec.2 and to quantitatively study the implication of $\delms$ measurements for  $B \to K \pi$ and $\phik$ decays. Finally in sec.4 we present our conclusions.

\section{Phase of $\bs$ Mixing}
The mass difference $\delmssm$ is calculated in the standard model from the \bsbsbar\ box
diagram, dominated by $t$-quark exchange:
\bea
M_{12}^s  & = & \frac{G_F^2}{6\pi^2}M_W^2M_{B_s}\left(\fbbs\right)
\hat{\eta}_{B_s} y_t f_2(y_t) \vert V_{ts}^*V_{tb}\vert^2 ,\nonumber\\
\delmssm  & = &|M_{12}^s|,\
\label{xs}
\eea
where $y_t=m_t^2/M_W^2$, and the function $f_2$ 
is the Inami-Lim function \cite{InamiLim}. $\hat{\eta}_{B_s}$ is the QCD correction and
$\fbbs $ represents the hadronization of the $B_s$ mixing operator in the SM.

New physics can generate operators of the type
\bea
\lnp \sim \sum \frac{c_{ij}}{ \Lambda^2} (\bar {s} \Gamma_i b) (\bar {s} \Gamma_j b),\
\label{NP}
\eea
that will  contribute to $ \delms$. Here $ \Lambda$ is the scale of new physics and $c_{ij}$ are the strength of the operators.  If the scale of NP is around a TeV the size $c_{ij}$ must be $ \sim |V_{ts}|^2$ to produce effects that can be of the same size as the experimental measurements \cite{model,zoltan, ball}. Experimental information on $\delms$ can constrain the $c_{ij}$ but this process is complicated by the fact that there can be more than one operator in Eq.~\ref{NP} and the fact that the different NP operators can hadronize differently than the SM operator. Assuming for the sake of simplification that only two operators contribute in Eq.~\ref{NP}, we can write
\bea
\ms & = & \mssm+ \msnp, \nonumber\\
\msnp & = & \frac{<\bs|\lnp|\bsbar>}{M_{B_s}}, \nonumber\\
\delms & = & \delmssm \left | 1 + r_1 \eta_1 e^{ i \phi_1}+r_2 \eta_2 e^{ i \phi_2} \right| ,\
\label{generalmixing}
\eea
where $r_1$ and $r_2$ represent the different hadronization of the NP operators relative to the SM operators. The quantities $\eta_{1,2}$ may be related to 
$c_{ij}$ in Eq.~\ref{NP} in a specific model of NP.
 One may combine to write the NP amplitudes as  a single complex number as
 \bea
\delms & = & \delmssm \left | 1 +  \eta e^{ i \phi} \right|, \nonumber\\
\eta e^{ i \phi} & = & r_1 \eta_1 e^{ i \phi_1}+r_2 \eta_2 e^{ i \phi_2} ,\
\label{generalmixing1}
\eea
where the relation between $\eta$ and the size of the NP operators is
complicated by SM uncertainties,  QCD effects of NP operators and more than one NP amplitude. 
As indicated in the introduction the measurement of the phase in $\bs$ mixing is a good place to look for NP.
{}From the general structure of Eq.~\ref{generalmixing1} it is trivially true that if $ \phi_{1,2}=0$ then there is no new NP weak phase in $\bs$ mixing. However, even with $\phi_{1,2} \ne 0$ it is possible that the effective phase $\phi$ in Eq.~\ref{generalmixing1} vanishes. This may be due to accidental cancellation or flavour symmetries
where the NP contributions may conspire to produce a small or no effective phase
$\phi$ even though $\phi_1$ and $\phi_2$ may be large.

 In this section we   study a flavour symmetry under which the new phase in $\bs$ mixing  remain hidden even with NP with nontrivial weak phases. We then study the implications for such a scenario for $\bsqq$ transitions.

The condition that the effective phase $\phi=0, \pi$ in Eq.~\ref{generalmixing1}
is
\bea
\msnp & = & {\msnp}^* 
 \Rightarrow  \nonumber\\ 
<\bs|\lnp|\bsbar> & = &
<\bsbar|(\lnp)^{\dagger}|\bs>. \
\label{npphase}
\eea
Since $CP|\bs> \sim |\bsbar>$ the condition in Eq.~\ref{npphase} is satisfied with the requirement
\bea
(\lnp)_{CP} & = & (\lnp)^{\dagger}, \
\label{cp}
\eea
where $(\lnp)_{CP}$ is just the CP transformed $\lnp$. However it is easy to see, and is well known, that the requirement in Eq.~\ref{cp} just leads to the trivial possibility that all weak phases
$\phi_{1,2}$ in Eq.~\ref{generalmixing}(Eq.~\ref{generalmixing1}) vanish. We will demonstrate this with two models of NP that we call vector-axial vector model (V/A) and the scalar-pseudoscalar model(S/P). We write the general NP lagrangian as
\bea
\lnpa & = & C_{LL,A} O_{LL,A} + C_{LR} O_{LR,A} + C_{RL,A} O_{RL}+ C_{RR, A} O_{RR, A},\nonumber\\
O_{LL,A} & = &  \bar{s} \gamma_A( 1-\gamma_5) b \bar{s}\gamma^A (1- \gamma_5) b, \nonumber\\
O_{LR,A} & = &  \bar{s} \gamma_A( 1-\gamma_5) b \bar{s}\gamma^A (1+ \gamma_5) b, \nonumber\\
O_{RL,A} & = &  \bar{s} \gamma_A( 1+\gamma_5) b \bar{s} \gamma^A(1- \gamma_5) b, \nonumber\\
O_{RR,A} & = &  \bar{s} \gamma_A( 1+\gamma_5) b \bar{s} \gamma^A(1+ \gamma_5) b, \
\label{vector-scalar}
\eea
where $A=S,V$  and $\gamma^A= 1, \gamma^{\mu}$ for the S/P and V/A NP models.
Now one easily checks that $(O_{X,Y})_{CP}= O_{X,Y}^{\dagger}$ where $X,Y=L,R$ for the operators in Eq.~\ref{vector-scalar} for both V/A and S/P models of NP. The condition in Eq.~\ref{cp} then leads to the fact that
the coefficients $C_{X,Y}$ in Eq.~\ref{vector-scalar} are real and we get the trivial scenario that all NP weak phases in $\lnp$ vanish.

However we show that the condition in Eq.~\ref{npphase} can also be satisfied with a flavour symmetry. Let us now define a flavour transformation, $T$, such that
\bea
T|s> & = & |b>, \nonumber\\
T|b> & = & |s>. \
\label{s-b}
\eea
This is nothing but the $s-b$ interchange transformation discussed in Ref~\cite{23} which is the quark equivalent of the $ \mu-\tau$ interchange symmetry used in the study of neutrino mixing \cite{mutau,moh}.
{}From Eq.~\ref{s-b} we obtain
\bea
T|\bs> & = & \xi|\bsbar> ,\nonumber\\
T|\bsbar> & = & \xi |\bs> , \nonumber\\
T \lnp T^{\dagger} & = & T  \left [\sum \frac{c_{ij}}{ \Lambda^2} (\bar {s} \Gamma_i b)(\bar {s} \Gamma_j b) \right]  T^{\dagger}, \nonumber\\
& = & \sum \frac{c_{ij}}{ \Lambda^2} (\bar {b} \Gamma_i s)(\bar {b} 
\Gamma_j s),  \
\label{s-bmeson}
\eea
where $|\xi|^2=1$.

Hence to recover the condition in Eq.~\ref{npphase} we demand that invariance of $\lnp$ under the $s-b$ flavour symmetry  implies
\bea
T \lnp T^{\dagger} 
& = & (\lnp)^{\dagger} \Rightarrow \nonumber\\
\sum \frac{c_{ij}}{ \Lambda^2} (\bar {b} \Gamma_i s)(\bar {b} \Gamma_j s)  & = &\sum_i \frac{c_{ij}^*}{ \Lambda^2} 
(\bar {b}\gamma_0 \Gamma_j^{\dagger} \gamma_0 s)(\bar {b}\gamma_0 \Gamma_i^{\dagger} \gamma_0 s) .\
\label{s-blag}
\eea 
Now writing $\gamma_0 \Gamma_{ij}^{\dagger} \gamma_0 = \pm \Gamma_{i'j'}$ the relation in Eq.~\ref{s-blag} then translates, in general, to
\bea
c_{ij}^*  & =  &  \pm c_{i'j'} \.
\label{coeff}
\eea
For the particular case of the V/A model of NP $\Gamma_{ij}=\gamma_{\mu}( 1 \pm \gamma_5)$ and we have
$\gamma_0 \Gamma_{ij}^{\dagger} \gamma_0 =  \Gamma_{ij}$ leading to
  $c_{ij}^*   =   c_{ij}$ and so the coefficient of each individual operator in $L_{NP,V}$ is real. This is easy to understand as in this case invariance under $s-b$ flavour symmetry is the same as requiring invariance under CP. It is possible to impose a weaker requirement for $s-b$ symmetry as
\bea
<\bsbar|T \lnp T^{\dagger}|\bs> 
& = & <\bsbar|(\lnp)^{\dagger}|\bs>. \
\label{s-blagweak}
\eea
This then leads to, using Eq.~\ref{vector-scalar},
\bea
<\bsbar|C_{LL,V} O_{LL,V}^{\dagger} + C_{LR} O_{LR,V}^{\dagger} + C_{RL,V} O_{RL}^{\dagger}+ C_{RR, V} O_{RR, V}^{\dagger}|\bs> =\nonumber\\
<\bsbar|C_{LL,V}^* O_{LL,V}^{\dagger} + C_{LR}^* O_{LR,V}^{\dagger} + C_{RL,V}^* O_{RL}^{\dagger}+ C_{RR, V}^* O_{RR, V}^{\dagger}|\bs>.\
\label{vector_relations}
\eea
Now as $ <\bsbar| O_{LL,V}^{\dagger}|\bs> = <\bsbar| O_{RR,V}^{\dagger}|\bs>$ and  $ <\bsbar| O_{LR,V}^{\dagger}|\bs> = <\bsbar| O_{RL,V}^{\dagger}|\bs>$ using parity transformation, we see that Eq.~\ref{s-blagweak} is also satisfied with the non trivial solution
\bea
C_{LL,V}^*=C_{RR,V}, \nonumber\\
C_{LR,V}^*=C_{RL,V}.\
\label{sb_vector}
\eea

{}For the S/P models $s-b$ symmetry of the lagrangian is possible with complex coefficients in the lagrangian  and here the relation in
  Eq.~\ref{coeff} can be written as,
\bea
C_{LL,S}^*=C_{RR,S}, \nonumber\\
C_{LR,S}^*=C_{RL,S}.\
\label{sb_scalar}
\eea
Hence $\lnpa$ has the form, using Eq.~\ref{sb_vector} and Eq.~\ref{sb_scalar} in Eq.~\ref{vector-scalar},
\bea
\lnpa & = & |a|\left [e^{-2i \phi}O_{LL,A} + e^{2i \phi} O_{RR,A}\right]+|b|
\left [e^{-2i \psi}O_{LR,A} + e^{2i \psi} O_{RL,A}\right].\
\label{lnps_final}
\eea
 We now ask the question if the phases $\phi$ and $\psi$ in Eq.~\ref{lnps_final} can produce CP odd effects in $\bsqq$ transitions.
 To answer  this question we consider  models of NP where the new interaction arises at tree level from the exchange of a heavy field.

 Assuming that $\lnpa$ in Eq.~\ref{vector-scalar} arises after integrating a heavy field then we can write
\bea
\lnpa & \sim J_{sb}^2, \nonumber\\
J_{sb,A} & = & \left[ \alpha \bar{s} \gamma_A( 1-\gamma_5) b + \beta \bar{s} \gamma_A( 1+\gamma_5) b \right] ,\
\label{vector-scalar_tree}
\eea
where $A=S,V$ and $\gamma^A= 1, \gamma^{\mu}$ for the S/P and V/A NP models.
The requirements from Eq.~\ref{sb_vector} and Eq.~\ref{sb_scalar} then allow us to write
\bea
J_{sb,A} & = & |\alpha| \left[e^{ i  \phi} \bar{s}\gamma_A ( 1+\gamma_5) b \pm e^{-i \phi} \bar{s} \gamma_A( 1-\gamma_5) b \right] .\
\label{sb_current}
\eea
The relative sign between the terms in Eq.~\ref{sb_current} would correspond to a scalar or a pseudoscalar exchange in the S/P model of new physics.
We will study such a model in details in the next section. 

We can now write Eq.~\ref{vector-scalar_tree} as
\bea
\lnpa & = & |\alpha|^2 \left [e^{-2i \phi}O_{LL,A} \pm  O_{LR,A} \pm  O_{RL,A}+ e^{2i \phi} O_{RR,A} \right] .\
\label{vector_scalar_tree_final}
\eea

Even though the phase $\phi$ in $\lnpa$ is not observable in $\bs$ mixing its effect may be observed in $\Delta B=1$, $ b \to s \bar{q} q$ transitions. We can write the effective Hamiltonian that generates the 
$\bsqq$ transition as
\bea
\heff & \sim & J_{sb} J_{qq}, \nonumber\\
J_{sb,A} & = & |\alpha| \left [e^{ i  \phi} \bar{s}\gamma_A ( 1+\gamma_5) b \pm  e^{-i \phi} \bar{s} \gamma_A( 1-\gamma_5) b \right], \nonumber\\
J_{qq,A} & = &  \left[ c_R \bar{q} \gamma^A( 1+\gamma_5) q \pm c_L^* \bar{q} \gamma^A( 1-\gamma_5) q \right].\
\label{nonleptonic}
\eea
Here $c_{L,R}$ are in general complex. We will first consider the case
$c_L=c_R=c$ as this relation is satisfied in the model to be considered in the next section.
Writing $c=|c|e^{ i \phi_c}$ we can rewrite $H_{eff}$ as
\bea
\heff & \sim & |\alpha||c|\left[e^{ i  \phi_{+}} H_{RR, A}  +e^{ -i  \phi_{+} }H_{LL,A} \pm e^{ i  \phi_{-}}H_{RL}  \pm e^{  -i \phi_{-}}H_{LR} \right], \nonumber\\
H_{LL,A} & = &  \bar{s} \gamma_A( 1-\gamma_5) b \bar{q}\gamma^A (1- \gamma_5) q, \nonumber\\
H_{RL,A} & = &  \bar{s} \gamma_A( 1+\gamma_5) b \bar{q} \gamma^A(1- \gamma_5) q, \nonumber\\
H_{LR,A} & = &  \bar{s}\gamma_A ( 1-\gamma_5) b \bar{q}\gamma^A (1+ \gamma_5) q,\nonumber\\
 H_{RR,A} & = &  \bar{s}\gamma_A ( 1+\gamma_5) b \bar{q}\gamma^A (1+ \gamma_5) q, \
\label{heffgeneral}
\eea
where $\phi_{\pm} = \phi \pm \phi_c$.

To demonstrate our point we can use factorization to calculate nonleptonic decays. Let us consider the decay $B^- \to K^- \pi^0$ and using the currents associated with the
relative negative sign in $J_{sb,qq}$, we can write
\bea
<K^- \pi^0|H_{eff}|B^-> & \sim & |\alpha||c|<K^-|e^{ i  \phi} \bar{s}\gamma_A ( 1+\gamma_5) b -  e^{-i \phi} \bar{s} \gamma_A( 1-\gamma_5) b|B^->
\nonumber\\
& & <\pi^0|e^{i \phi_c}
\bar{d}\gamma^A ( 1+\gamma_5) d - e^{-i \phi_c} \bar{d}\gamma^A ( 1-\gamma_5) d |0>,\nonumber\\
& \sim & 4|\alpha||c|e^{ i \pi/2 } \sin{\phi} \cos{\phi_c}
<K^-|\bar{s}\gamma_A b|B^-> \nonumber\\
& &<\pi^0|\bar{d} \gamma^A \gamma_5 d|0>.\
\label{kpi}
\eea
 We observe that an effective weak phase of $\pi/2$  appears in the $ B \to K \pi$ amplitude but this phase is unobservable if $\phi=0$. The effective weak phase may change due to non factorizable effects but the essential point here is that the weak phase $\phi$ can produce CP odd effects in nonleptonic $B$ decays while its effect is hidden in $\bs$ mixing. For the case of relative positive sign in $J_{sb,qq}$ the amplitude $ \sim \sin{\phi_c} \cos{\phi}$. Note that if a weak phase is only associated with flavour changing $s-b$ vertex then $\phi_c=0$ and there are no CP odd effects.

We can now look at the decay $\bd \to \phi K_s$. For this decay we have to fierz the operators in Eq.~\ref{heffgeneral} for the S/P model and
we can define the Fierzed operators,
\bea
H_{LL,S}^F & = & -\frac{1}{2N_c} \bar{q} ( 1-\gamma_5) b \bar{s} (1- \gamma_5) q 
-\frac{1}{8N_c} \bar{q}\sigma_{\mu \nu} ( 1-\gamma_5) b \bar{s}\sigma^{\mu\nu} (1- \gamma_5) q, 
\nonumber\\
H_{LR,S}^F & = &  -\frac{1}{2N_c}\bar{q}\gamma_{\mu} ( 1-\gamma_5) b 
\bar{s}\gamma^{\mu} (1+ \gamma_5) q, \nonumber\\
H_{RL,S}^F & = &  -\frac{1}{2N_c}\bar{q}\gamma_{\mu} ( 1+\gamma_5) b 
\bar{s}\gamma^{\mu} (1- \gamma_5) q, \nonumber\\
H_{RR,S}^F & = & -\frac{1}{2N_c} \bar{q} ( 1+\gamma_5) b \bar{s} (1+ \gamma_5) q 
-\frac{1}{8N_c} \bar{q}\sigma_{\mu\nu} ( 1+\gamma_5) b \bar{s}\sigma^{\mu \nu} (1+ \gamma_5) q, \
\label{fierzgeneral}
\eea
where $q=s$ and we have done also a color Fierz and dropped octet operators that do not contribute in factorization.

We can therefore write,
 \bea
 A(\bd \to \phi K_s) &= & A^{SM}+A^{NP}_{\pm LR \pm RL} + A^{NP}_{ LL + RR},\nonumber\\
 A^{SM} & = &-{G_{\sss F} \over \sqrt{2}}
 V_{tb}V_{ts}^* Z \left[ a_3^t+ a_4^t + a_5^t -\frac{1}{2}a_7^t
 -\frac{1}{2}a_9^t -\frac{1}{2}a_{10}^t \right. \nonumber\\ 
& & \hskip20truemm \left.  -a_3^c- a_4^c - a_5^c +\frac{1}{2}a_7^c
+\frac{1}{2}a_9^c +\frac{1}{2}a_{10}^c \right], \nonumber\\
A^{NP}_{\pm LR \pm RL} & \sim &  \pm <K_s| \bar{s}\gamma_{\mu} b |B>
 <\phi|\bar{s}\gamma^{\mu} s |0> \left[e^{i\phi_{-}} + e^{i\phi_{-}}\right],\nonumber\\
& \sim & \pm <K_s| \bar{s}\gamma_{\mu} b |B>
 <\phi|
\bar{s}\gamma^{\mu} s |0>\cos {\phi_{-}},
\nonumber\\ 
A^{NP}_{ LL + RR} & \sim &   <K_s| \bar{s}\sigma_{\mu \nu} b |B>
 <\phi|\bar{s}\sigma^{\mu \nu} s |0> \left[e^{i\phi_{+}} + e^{i\phi_{+}}\right],\nonumber\\
& \sim &
 <K_s| \bar{s}\sigma_{\mu \nu} b |B>
 <\phi|\bar{s}\sigma^{\mu \nu} s |0>\cos{ \phi_{+}}, 
\nonumber\\
Z &= & 2f_{\phi}m_{\phi}F_{BK}(m_\phi^2) \varepsilon^*\cdot p_B, \
\label{phiK}
\eea
where the SM contribution can be found in Ref.~\cite{dattarparity}. We therefore see that there are no new CP phases in $A(B_d \to \phi K_s)$. For the V/A models also it is easy to check that there are no NP phase in $\bd \to \phi K_s$. This follows simply from the fact that only vector currents contribute to this decay and hence the matrix element of all the  operators in $\heff$ are  equal. Explicitly,  consider the decay $\bd \to \phi K_s$ and using the currents associated with the
relative negative sign in $J_{sb,qq}$, we can write
\bea
<\phi K_s|H_{eff}|\bd> & \sim & |\alpha||c|<K_s|e^{ i  \phi} \bar{s}\gamma_{\mu} b -  e^{-i \phi} \bar{s} \gamma_{\mu} b|\bd>
\nonumber\\
& & <\phi|e^{i \phi_c}
\bar{s}\gamma^{\mu} s - e^{-i \phi_c} \bar{s}\gamma^{\mu} s |0>,\nonumber\\
& \sim & 4|\alpha||c|e^{ i \pi/2 } \sin{\phi}e^{ i \pi/2 } \sin{\phi_c}
<K^-|\bar{s}\gamma_A b|B^-> \nonumber\\
& &<\pi^0|\bar{d} \gamma^A \gamma_5 d|0>,\nonumber\\
& \sim & 4|\alpha||c|(-1) \sin{\phi} \sin{\phi_c}
<K^-|\bar{s}\gamma_A b|B^-> \nonumber\\
& &<\pi^0|\bar{d} \gamma^A \gamma_5 d|0>.\
\label{kpi}
\eea
This demonstrates that  there are no new CP phases in $A(\bd \to \phi K_s)$.

We now consider the case where $c_L \ne c_R$ in Eq.~\ref{nonleptonic}. {}For the case when $c_{L,R}$ are real  there is a new weak phase of $\pi/2$ in $\kpi$ for both the V/A and the S/P models. {}For the decay $\phik$ a new weak phase of $\pi/2$ arises in V/A models while for the S/P models the phase in this decay depends on the size of $c_{L,R}$. Moving on to the most general case with complex $c_{L,R}$ the new weak phase in $\kpi$ and $\phik$ will depend on the phase in the $ b\to s$ vertex as well as on the phases in $c_{L,R}$.

In the next section we will study a two higgs doublet model as an example
of a NP model with $s-b$ interchange flavour symmetry.

 \section {A Specific NP Model}
In Ref~\cite{23} we presented a 
 two Higgs doublet  model which has a 2-3
interchange flavour  symmetry in  the down  quark sector  like  the $\mu-\tau$
interchange symmetry in the leptonic  sector. The
 2-3  symmetry is assumed in the gauge basis where  the mass matrix has
off  diagonal terms and is fully  2-3 symmetric.   Diagonalizing  the mass
matrix  splits  the masses of $s$ and $b$ or  $\mu$ and $\tau$ and  leads to vanishing $m_s(m_\mu)$.
The breaking of the 2-3 symmetry is then introduced though the strange
quark  mass in  the quark  sector and  the muon  mass in  the leptonic
sector.  The  breaking of  the 2-3 symmetry  leads to  flavor changing
neutral currents  (FCNC) in  the quark sector  and the  charged lepton
sector that are suppressed by ${  m_s \over m_b}$ and ${ m_{\mu} \over
m_{\tau}}$ in addition to the mass  of the Higgs boson of
the second Higgs doublet. Additional  FCNC effects of similar size can
be generated  from the  breaking of the  $s-b$ symmetry in  the Yukawa
coupling of the second Higgs doublet. A full discussion of the model is given in Ref~\cite{23} and we will repeat some of the discussions here for completeness.

We consider a Lagrangian of the form,  
\beq  {\cal  L}^{Q}_{Y}=  Y^{U}_{ij} \bar  Q_{i,L}  \tilde\phi_1
U_{j,R}  +  Y^D_{ij}  \bar  Q_{i,L}\phi_1 D_{j,R}  +  S^{U}_{ij}  \bar
Q_{i,L}\tilde\phi_2 U_{j,R} +S^D_{ij}\bar Q_{i,L} \phi_2 D_{j,R} \,+\,
h.c. ,
\label{lag1}
\eeq
\noindent where $\phi_i$, for $i=1,2$, are the two scalar doublets of
a 2HDM,  while  $Y^{U,D}_{ij}$ and $S_{ij}^{U,D}$  are the non-diagonal
matrices of the Yukawa couplings. After diagonalizing the $Y$ matrix one can have FCNC couplings associated with the $S$ matrix.

For convenience  we   express $\phi_1$ and  $\phi_2$ in a
suitable basis  such that  only the $Y_{ij}^{U,D}$  couplings generate
the fermion masses. In such a basis one can write,
  \beq 
  \langle\phi_1\rangle=\left(
\begin{array}[]{c}
0\\ {v/\sqrt{2}}
\end{array}
\right)\,\,\,\, , \,\,\,\, \langle\phi_2\rangle=0 \,\,\,.  \eeq
\noindent The two Higgs doublets in this case are of the form,
\bea   
\phi_1   &=  &   \frac{1}{\sqrt{2}}\pmatrix{0   \cr  v+H^0}   +
 \frac{1}{\sqrt{2}}\pmatrix{ \sqrt{2} \chi^+ \cr i \chi^0}, \nonumber\\
 \phi_2 &= & \frac{1}{\sqrt{2}}\pmatrix{ \sqrt{2} H^+ \cr H^1+i H^2}. \
 \eea
 
In principle there  can be mixing among the neutral  Higgs but here we
neglect such mixing.
We assume  the doublet $\phi_1$  corresponds to the scalar  doublet of
the SM and  $H^0$ to the SM Higgs field.  In  addition, we assume that
the second  Higgs doublet does  not couple to the  up-type quarks($S^U
\equiv 0$). For the down type couplings in Eq.~\ref{lag1} we have,
\beq   {\cal  L}^{D}_{Y}=  Y^D_{ij}   \bar  Q_{i,L}\phi_1   D_{j,R}  +
S^D_{ij}\bar Q_{i,L} \phi_2 D_{j,R} \,+\, h.c.
\label{lag2}
\eeq We  assume the  following symmetries for  the matrices  $Y^D$ and
$S^D$:
\begin{itemize}

\item{ There  is a discrete symmetry under  which $d_{L,R} \rightarrow
-d_{L,R}$}

\item{ There is a $s-b$ interchange symmetry: $s \leftrightarrow b$}

\end{itemize}

The discrete symmetry involving the  down quark is enforced to prevent
$ s \to d $ transition because of constraints from the kaon system. It
also prevents  $ b \to d $  transitions since $B_d$ mixing  as well as
the value  of $ \sin { 2  \beta}$ measured in $\bd(t)  \to J/\psi \ks$
are consistent with  SM predictions. Although there may  still be room
for NP  in $ b \to d$  transitions, almost all deviations  from the SM have
been reported only in $ b \to s$ transitions and so we assume no NP in
$ b \to d$ transitions in this work.

The above symmetries then give  the following structure for the Yukawa
matrices,
\begin{eqnarray}
 Y^D &= &\pmatrix{y_{11} & 0 & 0  \cr 0 & y_{22} & y_{23}\cr 0 &y_{23}
& y_{22}}, \nonumber\\ 
S^D &= &\pmatrix{s_{11} & 0 & 0 \cr 0 & s_{22} &
s_{23}\cr 0 &s_{23} & s_{22}}. \
 \label{Yukawas}
\end{eqnarray}

The down quark  mass matrix,  $M^D$ is  now given by  $ M^D= {  v \over
\sqrt{2}}Y^D$.   The  matrix  $Y^D$  is  symmetric  and  choosing  the
elements in $Y^D$ to be real the mass matrix is diagonalized by,

\bea 
M^D_{diag} & = & U^T M^D U =\pmatrix{\frac{v}{\sqrt{2}}y_{11} & 0
&  0  \cr  0  &   \frac{v}{\sqrt{2}}(y_{22}-y_{23})  &  0\cr  0  &0  &
\frac{v}{\sqrt{2}}(y_{22}+y_{23})}, \nonumber\\ 
U &  = & \pmatrix{1 & 0
&   0  \cr   0   &  -\frac{1}{\sqrt{2}}   &  \frac{1}{\sqrt{2}}\cr   0
&\frac{1}{\sqrt{2}} &  \frac{1}{\sqrt{2}}}.\ 
\eea 
It is  clear that the
matrix $U$  will also diagonalize  the $S^D$ matrix when  we transform
the  quarks  from the  gauge  to the  mass  eigenstate  via $  d_{L,R}
\rightarrow U  d_{L,R}$. Hence there  are no FCNC effects involving  the Higgs
$\phi_2$.
 
The  down   quark  masses   are  given   by,
  \bea  m_d   &  =   &  \pm
\frac{v}{\sqrt{2}}y_{11},       \nonumber\\
  m_s       &=&      \pm
\frac{v}{\sqrt{2}}(y_{22}-y_{23}),   \nonumber\\
  m_b   &  =   &   \pm
\frac{v}{\sqrt{2}}(y_{22}+y_{23}).\ 
\eea

Since $m_s <<  m_b$ there has to be a  fine tuned cancellation between
$y_{22}$ and $y_{23}$ to produce  the strange quark mass. Hence, it is
more  natural to  consider  the symmetry  limit $y_{22}=y_{23}$  which
leads to  $m_s=0$. We then introduce  the strange quark  mass as a
small breaking of the $s-b$ symmetry and consider the structure,
\begin{eqnarray}
 Y^D_n &= &\pmatrix{y_{11} & 0 &  0 \cr 0 & y_{22}(1+2z) & y_{22}\cr 0
&y_{22} & y_{22}}, \
 \label{symbreak}
 \eea
with $z  \sim 2  m_s/m_b$ being a  small number.  Note that we  do not
break  the $s-b$ symmetry  in the  $2-3$ element  so that  the $Y^D_n$
matrix remains  symmetric. This down quark matrix  is now diagonalized
by,
\bea
M^D_{diag} & = & {W }^{T} M^D W 
 =\pmatrix{ \pm \frac{v}{\sqrt{2}}y_{11} & 0 &
0 \cr
0
&  \pm \frac{v}{\sqrt{2}}zy_{22} & 0\cr
0
&0 &
 \pm \frac{v}{\sqrt{2}}(2+z)y_{22}}, \nonumber\\
 W & = & 
 \pmatrix{1 & 0 &
0 \cr
0
& -\frac{1}{\sqrt{2}}(1-\frac{1}{2}z) & \frac{1}{\sqrt{2}}(1+\frac{1}{2}z)\cr
0
&\frac{1}{\sqrt{2}}(1+\frac{1}{2}z) &
 \frac{1}{\sqrt{2}}(1-\frac{1}{2}z)}.\
 \label{wmatrix}
 \eea
 In Eq.~\ref{wmatrix} we have dropped terms of $O(z^2)$.
 The down quark masses are now taken to be,
\bea
m_d & = &  \pm \frac{v}{\sqrt{2}}y_{11}, \nonumber\\
 m_s &=& \pm \frac{v}{\sqrt{2}}zy_{22}, \nonumber\\
m_b & = &  \pm \frac{v}{\sqrt{2}}(2+z)y_{22}.\
\eea
We find  $z \approx  \pm 2m_s/m_b$ and  now the transformation  to the
mass  eigenstate  will  generate   FCNC effects  involving  $  \phi_2$ which  is
proportional to $z \sim 2 m_s/m_b \sim \lambda^2$ where $\lambda$ is the
cosine of the Cabibbo angle. For
definiteness  we will  choose the  positive sign  for $z$  though both
signs are  allowed.
At this point we can consider two scenarios: the first   corresponds to the situation where the matrix $S^D$ is still $s-b$ symmetric. This will result in $\lnps$ with $s-b$ symmetry and hence no observable phase in $\bs$ mixing. In the second case we will allow for small breaking of the $s-b$ symmetry in $S^D$  in Eq.~\ref{Yukawas}. This will then result in $\lnps$ with broken $s-b$ symmetry and hence an observable phase in $\bs$ mixing.

{}For the first case,
   $S^D$  in the mass  eigenstate basis  has
the form,
\bea
S^D  \rightarrow  S^{D'} &= &\pmatrix{s_{d}e^{ i \phi_{dd}} & 0 &
0 \cr
0
& s_{s}e^{ i \phi_{ss}} & z s e^{ i \phi_{sb}} \cr
0
&z s e^{ i \phi_{sb}} &
 s_{b} e^{ i \phi_{bb}} }, \
 \label{sd_case_a}
\end{eqnarray}
where
 \bea
s_{d}e^{ i \phi_{dd}} & = & s_{11}, \nonumber\\
s_{s}e^{ i \phi_{ss}} & =& (s_{22}-s_{23} ), \nonumber\\
s_{b}e^{ i \phi_{bb}} & = & (s_{22}+s_{23} ),   \nonumber\\ 
{s}e^{ i \phi_{sb}} & = & s_{23}. \ 
\eea
 It is clear that  $S^{D'}$ above, is symmetric under $s-b$ interchange.

The Hamiltonian describing the interaction of the Higgs, $H^{1,2}$, is given by,
 \bea
H_S & = &  \frac{1}{ \sqrt{2}}\left[ S_{sb} \bar{s} ( 1+\gamma_5) b +S_{bs}\bar{b} ( 1+\gamma_5) s \right] H^1 +h.c, \nonumber\\
H_P & = & i\frac{1}{ \sqrt{2}}\left[ S_{sb} \bar{s} ( 1+\gamma_5) b +S_{bs}\bar{b} ( 1+\gamma_5) s \right] H^2 +h.c, \
\label{scalarint}
\eea
 The relevant currents, $J_{ij}$ are,
 \bea
 J_{sb}^{S(P)} & = &\frac{1}{ \sqrt{2}} \left[ S_{sb} \bar{s}(1+ \gamma_5)b \pm 
 S_{bs}^* \bar{s}(1- \gamma_5)b\right],\nonumber\\
 J_{qq}^{S(P)} & = &\frac{1}{ \sqrt{2}} \left[ S_{qq} \bar{q}(1+ \gamma_5)q  \pm 
S_{qq}^* \bar{q}(1- \gamma_5)q\right],\
 \label{current}
 \eea
 where the $\pm$ sign correspond to scalar or pseudoscalar interactions.
  Using Eq.~(\ref{sd_case_a}) we have,
  \bea
 S_{sb(bs)} & = &   s e^{ i \phi_{sb}}z ,\\
 S_{qq} & =& s_{q}  e^{ i \phi_{qq}}, \
 \label{matrix}
 \eea
where $q=d,s$.
After integrating out the heavy Higgs bosons, $H^{1,2}$, which we shall henceforth rename $S$ and $P$ bosons,  we can generate the following effective
Hamiltonian for $\Delta B=2$ and $\Delta B=1$ processes,
\bea
\lnp & = & \frac{1}{2 {m^2_{S(P)}}}z^2 s^2
\left[\pm e^{ i 2 \phi_{sb}}O_{RR}  \pm e^{ -i 2 \phi_{sb}}O_{LL} +O_{RL} +O_{LR} \right], \
\label{mixing}
\eea
where $O_{ab}$ are defined in Eq.~\ref{vector-scalar} with $\gamma_A=1$
and
\bea
\heff &= & 4\frac{G_F}{\sqrt{2}}z \frac{s s_{q}}{g^2}
\frac{m_W^2}{{m^2_{S(P)}}}
\left[\pm e^{ i  \phi_{+}}H_{RR}  \pm e^{- i  \phi_{+}}H_{LL}\right. \nonumber\\
&  & \left. +e^{ i  \phi_{-}}H_{RL} +e^{ -i  \phi_{-}}H_{LR} \right], \
\label{heff}
\eea
 where $\phi_{\pm}= \phi_{sb} \pm \phi_{qq}$, $H_{ab}$ are defined in Eq.~\ref{heffgeneral} with $\gamma_A=1$ and the $\pm$ sign is for $S$ and $P$ exchange interactions. Note that $\lnp$ has the same structure as in Eq.~\ref{vector_scalar_tree_final} and is symmetric under $s-b$ interchange.
 
 The expression for $\msnp$, obtained using the vacuum insertion approximation, is given as
 \bea
 \msnp & = & \frac{1}{2 {m^2_{S(P)}}}z^2 s^2
 \left[(4 \mp \cos{ 2 \phi_{sb}}\frac{10}{3})A +\frac{2}{3} \right]f_{B_s}^2 M_{B_s}, \nonumber\\
A & = &\frac{M_{B_s}^2}{(m_b+m_s)^2}. \
\label{bsmixing}
\eea
Here we assume that either the $S$ or $P$ interactions dominate and further we allow for a relative phase factor of $\pm 1$ between $\msnp$ and $\mssm$.
{}Following Ref.~\cite{ball} we will take the SM prediction for $\delmssm$ to be $\delmssm=23.4 ps^{-1}$ which then leads to $|1+ \eta|=0.74$(see Eq.~\ref{generalmixing1}) allowing for values of $|\eta_{min}|=0.26$ and $|\eta_{max}|=1.74$. These values can be converted to values for $s_{min} \approx 0.1 $ and $s_{max} \approx 0.25$ from Eq.~\ref{bsmixing} where we have chosen $\phi_{sb}= \pi/2$ for the pseudoscalar exchange interactions. This choice of $\phi_{sb}$ leads to the largest value for $s$. For our calculations we choose $m_s=100$ MeV, $m_b=4.8$ GeV, $f_{B_s}=250$MeV and $m_H=1$TeV. The choice for the higgs mass, $m_H$, is based on the fact that one expects new physics around $\sim $TeV to stabilize the standard model higgs mass. We assume that the same new physics needed for the higgs mass stabilization is also responsible for new FCNC effects in $B$ decays.

Let us now calculate the NP contribution to $ B^- \to K^- \pi^0$.
In $B\to\pi K$ decays, there are four classes of NP
operators, differing in their color structure: ${\bar s}_\alpha
\Gamma_i b_\alpha \, {\bar q}_\beta \Gamma_j q_\beta$ and ${\bar
s}_\alpha \Gamma_i b_\beta \, {\bar q}_\beta \Gamma_j q_\alpha$
($q=u,d$). The matrix elements of these operators are then
combined into single NP amplitudes, denoted by $\ApNPqph$ and
$\ApNPCqph$ respectively with $q=u,d$ \cite{BKpidecays2a, BKpidecays2b}.
 We have
\bea
<K^- \pi^0|\heff|B^-> & = & |{\cal A}^{\prime, comb}|e^{ i\Phi'}= 
 4\frac{G_F}{\sqrt{2}}\chi_s A_{dd} 
\left[\pm 2 i\sin{ \phi_{+}}- 2 i\sin{ \phi_{-}} \right], 
\nonumber\\
& = & 16
\frac{G_F}{\sqrt{2}}\chi_s A_{dd}e^{ i \pi/2}
\cases{\cos {\phi_{sb}} \sin {\phi_{dd}}  &
$H = S$ , \cr -\sin{\phi_{sb}} \cos {\phi_{dd}} &
$H = P$ , \cr}
\nonumber\\
\chi_{s} & = & \frac{2m_s}{m_b} \frac{ss_{d}}{g^2}
\frac{m_W^2}{m_{S(P)}^2},\nonumber\\
A_{dd} & = & <K^-| \bar{s} b|B^-><\pi^0| \bar{d} \gamma_5 d|0>.\
\label{acomb}
\eea
Here ${\cal A}^{\prime, comb}|e^{ i\Phi'} \equiv - \ApNPuph + \ApNPdph$.
In the case of scalar exchange the CP odd effect in the amplitude
 is due to the phase $\phi_{dd}$ while in the case of the pseudoscalar exchange the CP odd effect
 comes from the phase $\phi_{sb}$ which also appears in $\bs$ mixing. Hence we will consider the pseudoscalar exchange only for our quantitative analysis.

We first note that we can choose, in Eq.~\ref{acomb},  the phase $\Phi' = 90^0$ which is then consistent with a fit to the $ B \to K \pi$ data found in Ref.~\cite{BKpidecays2a} which  found $\Phi' \sim 100^0$ and 
 $| {\cal A}^{\prime, comb} / T' | = 1.64$ where $T'$ is the SM tree amplitude. Here we will attempt to see if this value for the ratio
 $| {\cal A}^{\prime, comb} / T' | $ is consistent with the $\delms$ measurements.
{}For the pseudoscalar exchange model we can write for the ratio
  $| {\cal A}^{\prime, comb} / T' | $ , using
naive factorization\cite{rhoKstar},
\bea
| {\cal A}^{\prime, comb} / T' |  \nonumber\\
&=&
\left \vert \frac{16 \sin{ \phi_{sb}} \cos { \phi_{dd}}\chi_{s}A_{dd}}
{(c_1+c_2/N_C)V_{ub}^*V_{us} \bra{\pi^0} {\bar b} \gamma_\mu (1 - \gamma_5) u \ket{B^-}
\bra{K^-} {\bar u} \gamma^\mu (1 - \gamma_5) s \ket{0} }
\right\vert,   
\nonumber\\
\label{constraint}
\eea
where $c_{1,2}$ are the Wilson's coefficients of the SM effective Hamiltonian. We can rewrite the above expression as,
\bea
| {\cal A}^{\prime, comb} / T' |  \nonumber\\
&=&
\left \vert {16 \sin{ \phi_{sb}} \cos { \phi_{dd}}\chi_{s}}
\right\vert M,   
\nonumber\\
M & = & \left \vert
\frac 
{
  [ ( m_{\sss B}^2 - m_{\sss K}^2) /
(m_b - m_s) ] F_0^{\sss K} ( m_\pi^2 / 2 m_d ) f_\pi / \sqrt{2} 
} 
{
 f_{\sss K} (m_{\sss B}^2 -
m_\pi^2) F_0^\pi / \sqrt{2}\left( c_1 + {c_2 \over N_c} \right) 
\left\vert V_{ub}^* V_{us}
\right\vert 
}
\right \vert \ 
\label{constraint1}
\eea
We take $(f_{\sss K}/f_\pi) (F_0^\pi/F_0^{\sss K}) \sim 1$, $|V_{ub}^*
V_{us}/V_{tb}^* V_{ts}| = 1/48$ and $c_1 + c_2/N_c = 1.018$. Now using the values of $s$ obtained from $\bs$ mixing we obtain the maximum possible values
$| {\cal A}^{\prime, comb} / T' | \approx (0.47, 1.2)s_d (6 MeV/m_d) $.
Assuming $s_d \sim 1$ in Eq.~\ref{acomb}, we find the lesser value for
$| {\cal A}^{\prime, comb} / T'|$  a bit smaller than the
   fit value in Ref.~\cite{BKpidecays2a} of 1.64.
  One could increase $| {\cal A}^{\prime, comb} / T'|$ by choosing $s_d$ large enough, $s_d \sim 3$, which would indicate a strongly coupled theory. The second, larger value for $| {\cal A}^{\prime, comb} / T'|$ is in better agreement with the fit in Ref.~\cite{BKpidecays2a}. One could also increase $| {\cal A}^{\prime, comb} / T'|$ by lowering $m_d$ which can be taken to be in the range 4-8 MeV \cite{pdg}.
Hence the predictions for the size of NP in $B \to K \pi$ decays does not rule out the scenario that there is no new phase in $\bs$ mixing. However a measurement of a new weak phase in $B_d \to \phi K_s$ would indicate new phase in $\bs$ mixing in this model.

To introduce a phase in $\bs$ mixing we have to break the $s-b$ interchange flavour symmetry. This can be done
by allowing $S^D$ in Eq.~\ref{Yukawas} to  break the $s - b$ interchange symmetry.
 We  choose,
\bea
 S^D &= &\pmatrix{s_{11} & 0 &
0 \cr
0
& s_{22} & s_{23}(1+ 2 \epsilon)\cr
0
&s_{23} &
 s_{22}(1+ 2 \delta)}, \
 \label{sd1g}
 \eea
 with $ \epsilon, \delta $ being  small quantities of the same size or smaller
 than $z$.  
 The transformation to the mass
 eigenstate leads to the general parametrization,
\bea
S^D  \rightarrow  S^{D'} &= & \pmatrix{s_{d}e^{ i \phi_{dd}} & 0 &
0 \cr
0
& s_{s}e^{ i \phi_{ss}} & zse^{ i \phi_{sb}} - \epsilon p  e^{ i \psi_{sb}}\cr
0
&zse^{ i \phi_{sb}}+ \epsilon p  e^{ i \psi_{sb}} &
 s_{b}e^{ i \phi_{bb}} }, \
 \label{sdm_sbbreak}
\end{eqnarray}
where
 \bea
s_{d}e^{ i \phi_{dd}} & = & s_{11}, \nonumber\\
s_{s}e^{ i \phi_{ss}} & =& (s_{22}-s_{23} + s_{22} \delta)+
s_{22}\delta z-s_{23} \epsilon, \nonumber\\
s_{b}e^{ i \phi_{bb}} & = & (s_{22}+s_{23} + s_{22} \delta)
-s_{22} \delta z + s_{23} \epsilon,   \nonumber\\ 
se^{ i \phi_{sb}} & = & s_{22}\frac{ \delta}{z} + 
s_{23}, \nonumber\\
pe^{ i \psi_{sb}}& = & s_{23}  . \ 
\label{sd1m}
\eea
There is now an additional FCNC involving $ b \to s $ transitions
whose source is the $s-b$ symmetry breaking in $S^D$. Note that the
2--3 off--diagonal elements in $S^{D'}$  contain a part that is symmetric under
$s-b$ interchange and a part that is antisymmetric under the $s-b$
interchange.  The parameters in $ S^{D'}$ can be obtained or
constrained from a fit to $B$ decay data. A phase in $\bs$ mixing arises from the interference of the the $s-b$ symmetric and the $s-b$ antisymmetric piece.

The structure in Eq.~\ref{sdm_sbbreak} is rather complicated and for the illustration of how a new phase in $\bs$ mixing may arise we can reduce the parameters in the model to simplify the discussion. Hence, we will assume  
 $ \delta=0$ which leads to $\phi_{sb}= \psi_{sb}, s=p $ and calling $s_{23} \equiv s$  we obtain,

\bea
S^D  \rightarrow  S^{D'} &= & \pmatrix{s_{d}e^{ i \phi_{dd}} & 0 &
0 \cr
0
& s_{s}e^{ i \phi_{ss}} & s (z- \epsilon ) e^{ i \phi_{sb}} \cr
0
&s(z + \epsilon ) e^{ i \phi_{sb}}  &
 s_{b}e^{ i \phi_{bb}} }, \
 \label{sdm_sb_break_simple}
\end{eqnarray}
The relevant currents, $J_{ij}$ are,
 \bea
 J_{sb}^{S(P)} & = &\frac{1}{ \sqrt{2}}s \left[ (z-\epsilon) e^{ i \phi_{sb}}
 \bar{s}(1+ \gamma_5)b \pm 
 (z+\epsilon) e^{ -i \phi_{sb}} \bar{s}(1- \gamma_5)b\right],\nonumber\\
 J_{qq}^{S(P)} & = &\frac{1}{ \sqrt{2}} \left[ S_{qq} \bar{q}(1+ \gamma_5)q  \pm 
S_{qq}^* \bar{q}(1- \gamma_5)q\right],\
 \label{current_new}
 \eea
 where the $\pm$ sign correspond to scalar or pseudoscalar interactions.
The $\bs$ mixing is now generated by,
\bea
\lnp & = & \frac{1}{2 {m^2_{S(P)}}}z^2 s^2
\left[\pm(1-\frac{\epsilon}{z})^2 e^{ i 2 \phi_{sb}}O_{RR}  
\pm(1+\frac{\epsilon}{z})^2 e^{ -i 2 \phi_{sb}}O_{LL} +(1-\frac{\epsilon^2}{z^2})(O_{RL} +O_{LR}) \right].\nonumber\\ 
\label{mixing_break}
\eea
Now we can try two approximations. One with ${\epsilon \over z}$ small and one with $ \epsilon \sim z$. 
 The possibility $\epsilon >> z$ is ruled out by the $B \to K \pi $ data as it does not produce a new weak phase \cite{23}.
 {}For ${\epsilon \over z}$ small, we get keeping only terms linear in
${\epsilon \over z}$,
\bea
\lnp & = & \nonumber\\ 
& & \hskip-2.5truecm \frac{1}{2 {m^2_{S(P)}}}z^2 s^2
\left[\pm  e^{ i 2 \phi_{sb}}O_{RR}  
\pm e^{ -i 2 \phi_{sb}}O_{LL} \pm \frac{2 \epsilon}{z} (e^{ -i 2 \phi_{sb}}O_{LL}  
- e^{ +i 2 \phi_{sb}}O_{RR}) 
 +   (O_{RL} +O_{LR}) \right]. \nonumber\\
\label{mixing_break_small}
\eea
We can now write $\msnp$ as,
\bea
 \msnp & = & \frac{2}{ {m^2_{S(P)}}}z^2 s^2
 \left[(1 \mp \frac{5}{6}\cos{ 2 \phi_{sb}})A +\frac{2}{12} \pm 
i\frac{5}{3}\frac{\epsilon}{z}\sin{ 2 \phi_{sb}}A
\right]f_{B_s}^2 M_{B_s}, \nonumber\\
A & = &\frac{M_{B_s}^2}{(m_b+m_s)^2}. \
\label{bsmixing_break}
\eea
So we see that there is an observable phase in $\bs$ mixing which is proportional to ${\epsilon \over z}$ and is given by,
\bea
\sin{ \phi_{B_s}} & = & \frac{Im M_{12}^{s,NP}}{|M_{12}^{s,NP}|},\nonumber\\
& \approx &\frac
{\frac{5}{3}\frac{\epsilon}{z}\sin{ 2 \phi_{sb}}A}
{(1 \mp \frac{5}{6}\cos{ 2 \phi_{sb}})A +\frac{2}{12}}.\
 \label{bsphase}
 \eea
We now study the predictions for $\bsqq$ transitions. To simplify the discussion we will choose $S_{qq}$ to be real and so 
any new phase in $\bsqq$ is due to the weak phase in $J_{sb}$. It is also clear that only the pseudoscalar exchange contribute to $\kpi$.
 One can check that there will be a new weak phase in $\bd \to \phi K_s$ proportional to ${\epsilon \over z}$ while the weak phase in $B^- \to K^- \pi^0$ will deviate from $\pi/2$ by an amount
proportional to ${\epsilon \over z}$. 

{}For $\epsilon=z$,
 The relevant currents, $J_{ij}$ are now given by,
 \bea
 J_{sb}^{S(P)} & = &  \pm \frac{1}{\sqrt{2}}s z e^{ -i \phi_{sb}}\bar{s}(1- \gamma_5)b,\nonumber\\
 J_{qq}^{S(P)} & = & 
 \frac{1}{\sqrt{2}}s_{q} 
 \left[ \bar{q}(1+ \gamma_5)q  \pm 
 \bar{q}(1- \gamma_5)q \right ],\
 \label{current_sbbreak}
 \eea
where again we have chosen $S_{qq}=s_q$ to be real.
It is now clear that $\lnp \sim J_{sb}^2$ has an observable phase which can be large and large NP phase can appear in $B_d \to \phi K_s$  and
 $B^- \to K^- \pi^0$. It is easy to check that the phase in $\kpi$ and $\phik$ is half the phase in $\bs$ mixing and hence if the $\bs$ mixing is observed to be small it would indicate a small new weak phase in $\kpi$ and in $\phik$ in contrast to the case considered above. In passing we note that if $\phi_{sb}= \pi$ then a NP phase of $\pi/2$ can appear in $\kpi$ but a  phase will also appear in the decay $\phik$ in contrast to the $s-b$ symmetric case considered above.

To summarize, if the $\bs$ mixing phase is observed to be small and is due to the small breaking of the $s-b$ interchange symmetry then one should observe a large new CP odd phase, close to $\pi/2$, in the decay $\kpi$ and a small new CP odd phase in $\phik$. On the other hand if the
$\bs$ mixing is small due to small NP phase in the $ b \to s $ current then we should observe a small NP phase in the decay $\kpi$ and in $\phik$.

\section{Conclusions}
 In summary, in this paper we have studied the implication of the $\delms$ measurement on $\bsqq$ transitions. We pointed out that in a fairly general class of NP the new physics phase of $\bs$ mixing may be unobservable if there is a $s-b$ interchange flavour symmetry even in the presence of new CP odd phase in the $ b \to s $ transition. This phase can however appear in certain $\bsqq$ transitions like $ B \to K \pi$ decays but not in others like the $\bd \to \phi K_s$ decay. Working within a two higgs doublet model we calculated the allowed NP contribution to $B \to K \pi$ decays  with the new $\delms$ measurement. We also showed that if the NP weak phase is only in the $ b \to s$ current and a small observed $\bs$ mixing phase is due to the small breaking of the $s-b$ interchange symmetry then one should observe a large new CP odd phase, close to $\pi/2$, in the decay $\kpi$ but a small CP odd phase in $\phik$. On the other hand a small $\bs$ mixing phase and  a small NP phase in the decays $\kpi$ and $\phik$
 would indicate the absence of the $s-b$ flavour symmetry.
 
\bigskip
\noindent {\bf Acknowledgements}:
This work is financially supported by NSERC of Canada. We thank  R.N. Mohapatra  for useful discussion.


\end{document}